# Slip electron flow in GaAs microscale constrictions

D. I. Sarypov[1,2,*], D. A. Pokhabov[1,2], A. G. Pogosov[1,2], E. Yu. Zhdanov[1,2], A. A. Shevyrin[1],

A. A. Shklyaev[1,2], A. K. Bakarov[1,2]

[1]Rzhanov Institute of Semiconductor Physics, Siberian Branch, Russian Academy of Sciences, Novosibirsk, 630090, Russia

[2]Novosibirsk State University, Novosibirsk, 630090, Russia

* corresponding author: d.sarypov@g.nsu.ru

**Hydrodynamic electron transport in solids, governed by momentum-conserving electron-electron collisions, offers a unique framework to explore collective phenomena. Within this framework, correlated electron motion is modeled as viscous fluid flow, with viscosity serving as the interaction parameter. Advances in electron hydrodynamics remain constrained by two unresolved issues: the questionable existence of perfect boundary slip—a hallmark of frictionless transport—in electron fluids, and the lack of quantitative experimental confirmation of the theoretical relation linking the viscosity to electron-electron scattering length. Here, we resolve this through independent measurements of these quantities in the same electron system in GaAs/AlGaAs heterostructure. Our experiments provide direct evidence of perfect boundary slip in microscale constrictions—unprecedented phenomenon for electron liquid that parallels ultrafast water transport in carbon nanotubes. These findings bridge the fields of electron hydrodynamics and nanofluidics, highlighting the transformative potential of hydrodynamic engineering across condensed matter and fluidic technologies.**

In sufficiently pure electron systems, such as a two-dimensional electron gas (2DEG) in graphene and GaAs, frequent momentum-conserving electron-electron (e-e) collisions can induce collective hydrodynamic behavior, mirroring phenomena in classical fluids[1-4]. In such conditions, the hydrodynamic description of electron transport becomes a powerful tool, that enables the discovery of striking many-body effects, including the Gurzhi resistance reduction[5-6], electron whirlpools[7-10], Poiseuille-like flow in confined channels[11-14], superballistic transport[15-18] and viscous Hall-effect[19,20]. Central to these phenomena is the electron viscosity, a key parameter encoding e-e interaction strength and setting the hydrodynamic onset conditions.

The hydrodynamic behavior of electrons in solids is fundamentally governed by boundary conditions—a factor that remains intensely debated in electron hydrodynamics. While conventional theories assume no-slip boundaries to explain phenomena like Poiseuille flow in graphene channels[11-13], recent studies of water transport in micro- and nanofluidic systems[21-27] have revealed the presence of boundary slip. Whether such slip occurs in electron fluids, however, remains unresolved, with conflicting theoretical predictions[28,29] and no direct experimental evidence. Equally critical is the lack of experimental data quantitatively verifying the conventional kinetic theory relation $\nu = \frac{1}{4}v_{\rm F} l_{\rm ee}$, which posits identical temperature ($T$) dependencies for viscosity $\nu$ and e-e scattering length $l_{\rm ee}$. Strikingly, recent measurements of viscosity[16,30,31] and e-e scattering length[32-35] reveal divergent $T$-dependences, challenging established models and underscoring the need for systematic experimental validation.

In this paper, we address these challenges through independent $\nu$ and $l_{\rm ee}$ measurements in a single device based on GaAs/AlGaAs 2DEG. This unified experimental approach eliminates ambiguities from sample-to-sample variations and makes it possible to directly probe the relation between viscosity and e-e scattering length. Our measurements not only confirm the low-$T$ prediction of kinetic theory but also reveal perfect boundary slip in electron flow. This slippage—analogous to frictionless water transport in carbon nanotubes[22-27]—distinguishes GaAs devices from graphene systems, where rough edges induce Poiseuille flow. While the carbon nanostructures are at the forefront of nanofluidics[36,37], our findings make GaAs systems with slippery boundaries preferable for low-dissipation electronic applications.

## Results

**Nanostructures for viscosity and scattering length measurements.** To measure the viscosity and e-e scattering length, we fabricated Hall bars on the basis of GaAs/AlGaAs heterostructure with high-mobility 2DEG. The studied Hall bars contain point contacts (PCs) playing a role of viscosimeter, and a magnetic focusing device for measurement of $l_{ee}$ (Figs. 1a and 1b). A feature of our heterostructure is the presence of the sacrificial $Al_{0.8}Ga_{0.2}As$ layer under 2DEG, which can be removed in order to detach the structure from the substrate, thereby create a freely suspended device (Fig. 1c) and access the unique experimental conditions of enhanced e-e interaction[38]. The suspension was recently found out to reduce viscosity[30] and e-e scattering length[35]. It is used in this work as an additional method to verify the relation between $\nu$ and $l_{ee}$. The studied 2DEG has the low-T concentration and mobility of $n = 6.5 \cdot 10^{11}$ cm-2 and $\mu \approx 2 \cdot 10^6$ cm2/(Vs), respectively. At the given $n$, the Fermi energy is $E_F \approx 23$ meV ($E_F/k_B \approx 270$ K). We note, that the suspension process does not ruin the quality of our structures and leaves concentration and mobility almost unchanged (Supplementary Note 1). A high electron mobility in the created GaAs 2DEG makes it possible to neglect the effects of momentum-relaxing (electron-impurity and electron-phonon) collisions in the temperature range up to 50 K (Fig. 1d). Details of device fabrication and transport properties are given in "Methods" section and Supplementary Note 1.

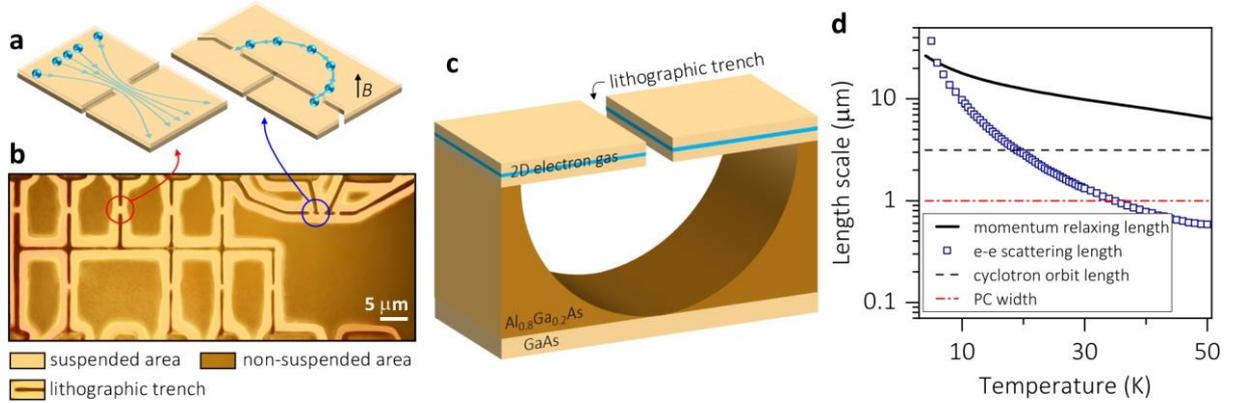

**Figure 1| Devices for measurements of electron viscosity and e-e scattering length in GaAs/AlGaAs heterostructure**. **a**, Schematic view of the point contact playing a role of viscosimeter and magnetic focusing device for measurement of e-e scattering length. **b**, Optical micrograph of one of the samples. Color coding denotes non-suspended, suspended areas and lithographic trenches. **c**, Heterostructure GaAs/AlGaAs with the sacrificial $Al_{0.8}Ga_{0.2}As$ layer, which can be selectively removed to suspend the structures. **d**, Key length scales for non-suspended electron system. Solid line is momentum relaxation length obtained from longitudinal resistance measurements (Supplementary Note 1), squares denote e-e scattering length measured below by transverse magnetic focusing technique, dashed and dash-dotted lines correspond to resonant cyclotron orbit length in magnetic focusing experiments (3.1 μm) and characteristic point contact width (1 μm), respectively.

The viscosity measurement method is based on the effect of superballistic conductance, which manifests itself as a non-monotonic *T*-dependence of the PC resistance[16]. A direct relation between the superballistic conductance and the electron viscosity[15] makes it possible to extract the viscosity values from the experiment. The superballistic conductance has a dependence on the PC width[15], distinctive from the conventional ballistic conductance[39,40], so we fabricated PCs with width varying between 0.9 μm and 2.0 μm (Fig. 1b) to distinguish the superballistic conductance and to test the validity of the viscosity measurements.

To precisely measure the e-e scattering length, we employed the transverse magnetic focusing technique—a well-established method for probing Fermi surface properties and many-body effects in solids[39,41-47]. Typical magnetic focusing device consists of two parallel PCs, that act as injector and detector of ballistic electrons (Fig. 1a). In our case the injector and detector have a width of 1 μm and placed at a distance of 2 μm from each other. The technique implies the use of weak transverse magnetic field $B$, that forces electrons to move along cyclotron trajectories with the diameter of $d_c = 2mv_F/(eB)$ ($m$ is effective electron mass). When the cyclotron diameter coincides with the injector-detector distance, a geometric resonance occurs, corresponding to the injected electrons hitting the detector, as illustrated in Fig. 1a. In experiment, it is usually observed as a focusing peak in the detector signal[42]. The sensitivity of the geometric resonance to e-e scattering provides a way to precisely measure the e-e scattering length as a parameter of thermal suppression of the focusing peak[32,35].

Combining the magnetic focusing devices and superballistic PCs in a single Hall bar, we can measure the e-e scattering length and electron viscosity independently and investigate the relation between them using transport measurements only.

**Measurements of e-e scattering length.** Figure 2a shows signal on detector $V_{\text{det}}$ normalized by injector current $I_{\text{inj}}$ as a function of magnetic field. It is measured using the scheme illustrated in the inset in Fig. 2b. A distinct peak is observed at magnetic field of $B^* = 0.13$ T. The cyclotron diameter $d_c^* = 2mv_F/(eB^*) = 2$ μm, corresponding to $B^*$, matches the injector-detector distance and meets the focusing condition. One can note that suspension of the device does not alter the position of the focusing peak (Fig. 2a), indicating that the suspension process leaves the electron density unchanged.

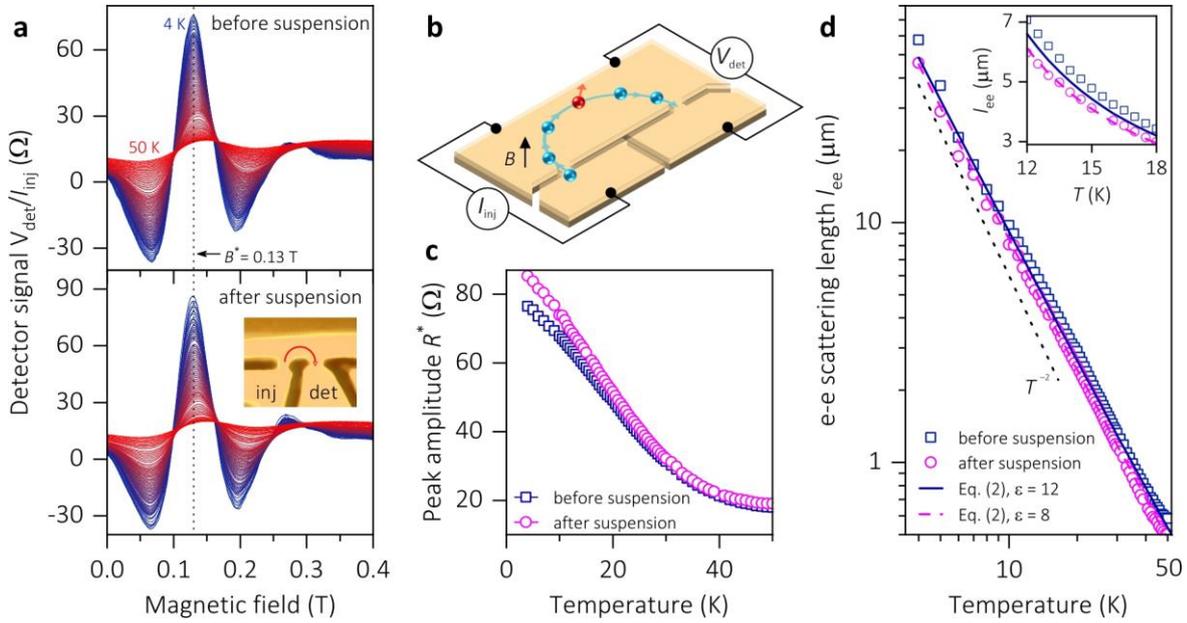

**Figure 2| Measurement of electron-electron scattering length using magnetic focusing technique**. **a**, Detector signal as a function of transverse magnetic field, measured before (top panel) and after (bottom panel) the suspension at different temperatures from 4 K (blue curve) to 50 K (red curve). The electron focusing occurs at magnetic field $B^* = 0.13$ T, marked by thin dotted line. Inset shows optical micrograph of suspended focusing device, red arrow shows resonant ballistic trajectory with the diameter of $d_c^* = 2$ μm. **b**, Measurement scheme for magnetic focusing. Red ball illustrates e-e scattering event destroying the resonant trajectory. **c**, The value of the detector signal at $B^*$, suppressing with increasing temperature. **d**, The values of e-e scattering length $l_{\text{ee}}$ measured by magnetic focusing before and after the suspension. Lines show the theoretical $l_{\text{ee}}$ values obtained with Eq. (2). Inset shows the same data, but on enlarged scale.

As the temperature increases, the focusing peak is suppressed. In the studied temperature range, the momentum relaxation length significantly exceeds the length of the resonant trajectory (Fig. 1d), so the observed suppression of the focusing peak is attributable to e-e scattering. The e-e scattering destroys the resonant ballistic trajectory, connecting the injector and detector at $B = B^*$, as schematically shown in Fig. 2b. To quantify the influence of e-e scattering on the resonant trajectory of electrons, we plot the value of the detector signal at $B^*$ as a function of temperature (Fig. 2c). Note that suspension of the device leads to greater suppression of the focusing peak due to enhanced e-e interaction in suspended samples[38].

The thermal suppression of the peak amplitude $R^*(T)$ can be described as follows[32,35]:

$$R^*(T) = R_\infty + R_0 \exp\left(-\frac{\alpha l}{l_{ee}(T)}\right). \tag{1}$$

Here $R_\infty$ and $R_0$ are the parameters describing high-$T$ and low-$T$ values of $R^*$, $l = \pi d_c^*/2$ is the length of the resonant trajectory and the parameter $\alpha$ accounts the effect of small-angle scattering and depends on the geometry of the magnetic focusing device[48,49]. The second term in Eq. (1) denotes the electron path length distribution with mean free path of $l_{ee}$.

We fit the data in Fig. 2c using Eq. (1) to extract $l_{ee}$ (Supplementary Note 2). The latter is shown in Fig. 2d as a function of temperature. One can see that $l_{ee}$ decreases after suspension. This means that e-e scattering probability $W_{ee} \propto 1/l_{ee}$ is increased in suspended samples. To describe this change, we employ the commonly used formula for $l_{ee}$ in 2D electron systems[50,51]:

$$l_{ee} = \frac{4\pi}{k_F}\left(\frac{E_F}{k_B T}\right)^2 \left[\ln\left(\frac{E_F}{k_B T}\right) + \ln\left(\frac{2q_{TF}}{k_F}\right) + 1\right]^{-1}, \tag{2}$$

where $k_F$ is the Fermi wavenumber and $1/q_{TF} = \varepsilon^* \hbar^2/(2me^2)$ is the screening length depending on the effective dielectric constant $\varepsilon^*$. The theory behind Eq. (2) describes the observed reduction of $l_{ee}$ by a decrease of the effective dielectric constant $\varepsilon^*$ (1.5 times after suspension), which is consistent with previous measurements[35]. The overall quantitative agreement of our measurements with the common theory (Fig. 2d) makes the measured $l_{ee}$ values a benchmark to compare with the viscosity.

**Measurements of electron viscosity.** As a next step, we extract the electron viscosity from the measurements of superballistic conductance[16]. We observe this effect as a fall of the PC resistance $R$ below the ballistic limit, as temperature increases (Fig. 3a). The resistance $R$ starts to grow at temperature above 50 K, which is expected since the momentum relaxing length becomes comparable with the PC width (Fig. 1d) due to rising electron-phonon scattering rate. To subtract the contribution of momentum-relaxing collisions to resistance $R$ we use an auxiliary sample without PC. The resistance $R_{2DEG}(T)$ of the auxiliary sample growths monotonously with $T$, showing no effect of superballistic conductance (Fig. 3b). By combining the measured $R(T)$ and $R_{2DEG}(T)$ dependences, we directly extract the superballistic conductance without any numerical calculations and fitting parameters (Supplementary Note 3).

Further we address our experiment to the problem of viscous electron flow through PC, which has a closed-form solution for superballistic conductance $G_{vis}$[16]:

$$G_{vis} = \frac{\pi}{16}\frac{e^2}{h}\frac{k_F v_F w^2}{\beta v} \tag{3}$$

Here $w$ is the PC width. Notably, the no-slip and perfect slip boundary conditions results in $G_{vis}$ values that differ twofold[28,52], which we parametrize by the factor $\beta$ (taking the values of 2 or 1 in case of no-slip or perfect slip boundary conditions, respectively). The question of friction at PC edges can be unambiguously resolved by comparing the $G_{vis}$ values with measured $l_{ee}$, as we show below.

In spite of unclear boundary conditions, Eq. (3) allows us to check the $w$-scaling of $G_{vis}$ and ensure, that the $G_{vis}$ values are correctly extracted from the experiment (Supplementary Note 3). To illustrate the change in the viscosity value after suspension we plot the quantity $G_{vis}/w^2$ (Fig. 3c). One can see that the superballistic conductance $G_{vis}$ enhances after suspension. Since the suspension leaves electron concentration and mobility almost unchanged (Supplementary Note 1), the observed enhancement of $G_{vis}$ is attributed solely to the change in the viscosity.

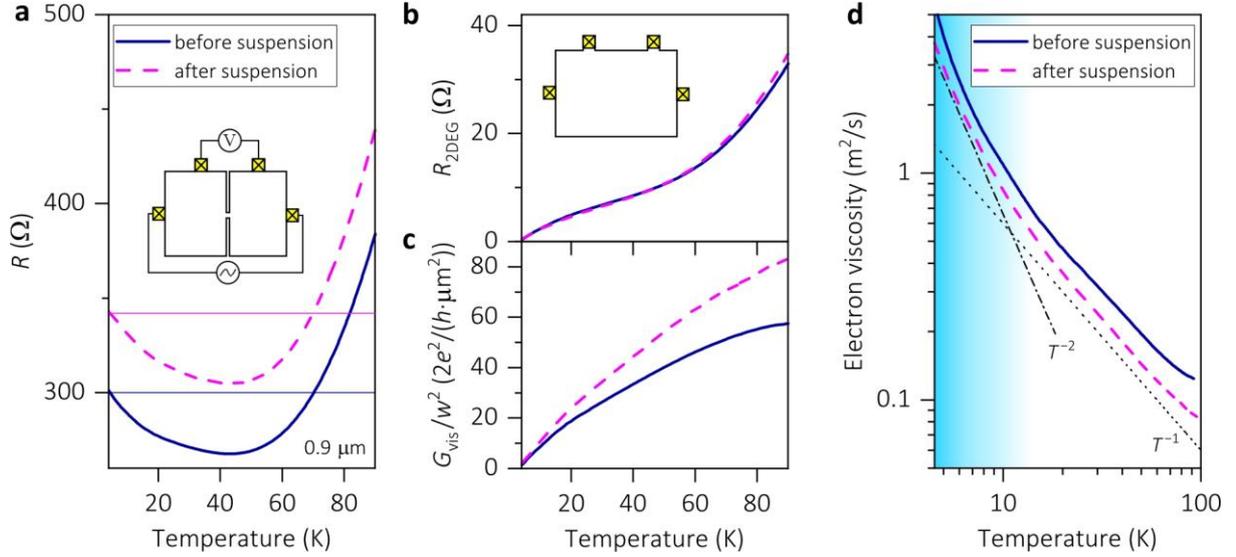

**Figure 3 | Extraction of the electron viscosity from the effect of superballistic conductance. a**, The resistance $R$ as a function of temperature for the PC with the lithographic width of 0.9 μm. Solid (dashed) line shows measurements performed before (after) the suspension according to the scheme shown in inset. Thin horizontal lines mark the values of low-$T$ ballistic resistance. **b**, The same data, but for reference sample with resistance $R_{2DEG}$ containing no PC inside. **c**, $T$-dependence of superballistic conductance $G_{vis}$, normalized by the square of the effective PC width $w^2$. **d**, $T$-dependence of the electron viscosity $\beta\nu$, obtained from Eq. (3), on a log-log scale. Dash-dotted and dotted lines denote $T^{-2}$ and $T^{-1}$ dependence, respectively. Shaded region marks the range of low temperatures ($k_B T \lesssim 0.05 E_F$). All panels have the same line stroke type and color coding.

As the $G_{vis}(T)$ dependence is extracted, we can convert it to the viscosity using Eq. (3). Before analyzing boundary conditions, we first discuss the $T$-dependence of viscosity $\beta\nu(T)$, which is shown in Fig. 3d. At low temperatures compared with the Fermi energy ($k_B T \ll E_F$, shaded region in Fig. 3d), we observe the $\nu \propto 1/T^2$ dependence which is consistent with Fermi liquid theory and measurements of $l_{ee}$ (Fig. 2d), while, at moderate temperatures, one can see a different $T$-dependence $\nu \propto 1/T$, in contrast to $l_{ee} \propto 1/T^2$ (Fig. 2d). This complex relation between $\nu$ and $l_{ee}$ was recently attributed to the thermal smearing of the Fermi surface[30]. After that, the suspension of the samples leads to noticeable reduction of the viscosity, similar to the suppression of e-e scattering length, which is also caused by enhancement of the e-e interaction in suspended structures.

The observed $\nu \propto 1/T$ dependence at moderate temperatures (Fig. 3d) invites a comparison to tomographic dynamics of electron fluid[53], a regime where collinear and head-on electron collisions cause different relaxation times of collective excitations[54,55]. While this regime has unconventional features observed in our experiment—such as $\nu \propto 1/T$ dependence[56] and the absence of Knudsen minimum[57]—our data reveal inconsistencies with tomographic model. First, the low-$T$ viscosity follows the Fermi-liquid $\nu \propto 1/T^2$ behavior (Fig. 3d), while the tomographic model predicts the $\nu \propto 1/T$ trend to persist down to the lowest temperatures[56]. Second, the superballistic conductance exhibits conventional $w^2$ scaling, distinct from fractional $w^{5/3}$ power law inherent to tomographic regime[56], at least, at moderate temperatures (Supplementary Fig. 3c). At the same time, measurement precision limits our ability to distinguish $w^2$ and $w^{5/3}$ scaling at $k_B T \ll E_F$ (Supplementary Fig. 3d). Given these inconsistencies, we cannot strictly ascribe the $\nu \propto 1/T$ dependence in our experiment to tomographic dynamics.

**Fundamental relation between viscosity and e-e scattering length.** By combining independent measurements of superballistic conductance $G_{vis}$, which can be converted to electron viscosity $\nu$, and the e-e scattering length $l_{ee}$, we directly test the kinetic theory prediction $\nu = \frac{1}{4} v_F l_{ee}$. This approach resolves ambiguity in boundary condition selection in Eq. (3). Using these experimental constraints, we refine Eq. (3) to incorporate both measured quantities:

$$\frac{\nu}{\frac{1}{4}v_\text{F} l_\text{ee}} = \frac{1}{\beta} \times \frac{\pi}{4}\left(\frac{e^2/h}{G_\text{vis}}\right)\frac{k_\text{F} w^2}{l_\text{ee}}. \quad (4)$$

Figure 4 shows the relation $4\nu/v_\text{F} l_\text{ee}$, defined via Eq. (4), as a function of temperature. Below 15 K, this relation saturates to a constant value, meaning that $\nu$ and $l_\text{ee}$ have the same $T$-dependence and hence proportional to each other. Suspending the samples yields analogous behavior, indicating that dielectric environment modifications equally impact low-$T$ $\nu$ and $l_\text{ee}$ values, thereby reinforcing measurement consistency. Critically, analysis of boundary conditions reveals stark contrasts: perfect slip scenario ($\beta = 1$) aligns with the conventional kinetic theory prediction, whereas no-slip model ($\beta = 2$) deviates significantly from experimental data. This experimental dichotomy clearly demonstrates slip electron flow through GaAs point contacts.

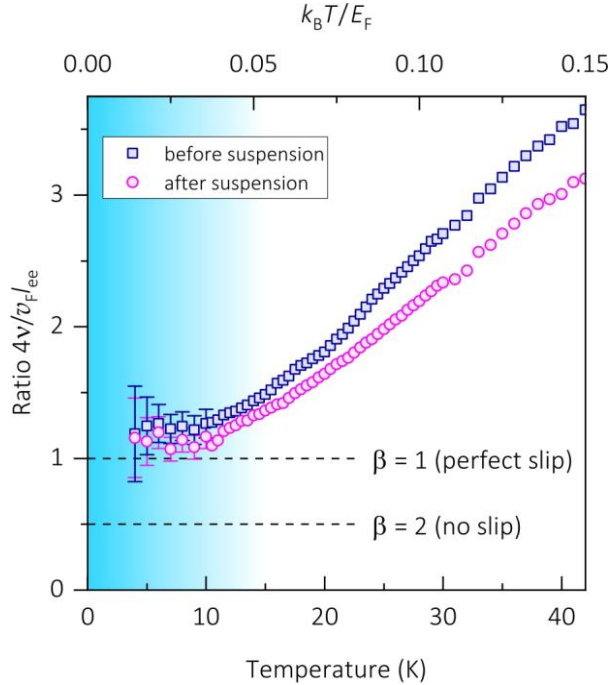

**Figure 4| The ratio of electron viscosity to electron-electron scattering length.** Temperature dependence of the ratio $4\nu/v_\text{F} l_\text{ee}$, with the electron viscosity $\nu$ and e-e scattering length $l_\text{ee}$ measured independently. Error bars at low temperature are estimated from reproducibility of the viscosity measurements (Supplementary Note 4). Shaded region marks the range of low temperatures ($k_\text{B} T \lesssim 0.05 E_\text{F}$). Dashed horizontal lines mark low-$T$ values of $4\nu/v_\text{F} l_\text{ee}$ ratio in case of perfect slip ($\beta = 1$) and no-slip ($\beta = 2$) at PC boundaries.

Rise in temperature above 15 K leads to the increase of the $\nu/l_{ee}$ ratio, reflecting the transition in $T$-dependence of viscosity. Namely, at $T > 15$ K, viscosity follows the $\nu \propto 1/T$ law (Fig. 3d) due to Fermi surface smearing, while e-e scattering length scales as $l_\text{ee} \propto 1/T^2$ (Fig. 2d), yielding $\nu/l_\text{ee} \propto T$ dependence. Moreover, we see the divergence between the data obtained before and after the suspension, that reveals a much stronger effect of suspension on the viscosity value compared to that of e-e scattering length. This divergence supplements recent claims[16,30] about non-trivial relation between $\nu$ and $l_\text{ee}$ at moderate temperatures, necessitating advanced theoretical approach to describe the viscosity.

**Discussion**

To summarize, direct measurements of the relationship between electron viscosity and e-e scattering length within a single electron system in GaAs confirm the coupling of these parameters at sufficiently low temperatures. Strikingly, we observe frictionless electron flow through microscale point contacts — a phenomenon that initially appears counterintuitive, as viscous behavior in conventional hydrodynamics arises from the momentum exchange between fluid elements and boundaries, typically leading to a Poiseuille-like flow profile. However, the viscous effects in our system emerge not within the PC itself, but in the adjacent regions. These effects cause electron drag towards the PC opening, which enhances conductance by redirecting the carriers, that do not contribute the conductance in ballistic regime[58]. The slippery PC boundaries amplify this drag by minimizing momentum loss at the edges, resulting in a conductance exceeding the predictions for no-slip boundaries (Eq. (3)).

This behavior parallels the ultrafast water transport observed in carbon nanotubes[22-27], where molecular slip at atomically smooth walls drastically enhances the flow rate. While the microscopic mechanisms differ—electron slip here originates from edge depletion, which suppresses boundary roughness (Supplementary Note 1), rather than molecular interactions[59]—both systems display how boundary conditions can fundamentally reshape hydrodynamic transport. Notably, the observed boundary slip is intrinsic to GaAs microconstrictions fabricated via etched trenches, requiring no external modifications (e.g., electrostatic gating[31]) to achieve dissipationless flow. By bridging solid-state electron hydrodynamics and nanofluidics, our work positions GaAs devices with smooth edges as a platform for energy-efficient electronic operation for future technological application.

Looking forward, we propose the suspended semiconductor structures as a promising system for exploring nonlinear hydrodynamic phenomena in solids, where low electron viscosity is critical. Namely, device suspension strongly reduces the viscosity (up to 20 % stronger than e-e scattering length) by enhancing e-e interaction and enabling access to high-Reynolds-number regimes. This lowers the threshold for observing turbulent electron flow that remains experimentally elusive but is theoretically predicted to emerge in solids[60,61].

**Methods**

**Growth of heterostructure.** Experimental samples are fabricated on the basis of GaAs/AlGaAs heterostructures grown by molecular beam epitaxy on semi-insulating (001) GaAs substrate. At first, a sacrificial $Al_{0.8}Ga_{0.2}As$ layer with a thickness of 400 nm is grown on the substrate. Then, it is covered by the GaAs/AlGaAs heterostructure with a thickness of 166 nm, which is a superlattice consisting of alternating layers of GaAs (2.3 nm thick) and AlAs (1.15 nm thick). The heterostructure contains 2DEG in the GaAs layer with a width of 13 nm, which is a symmetric quantum well for electrons, lying at a depth of 90 nm under the surface. Doping Si δ-layers with a density of $1.5 \cdot 10^{12}$ cm$^{-2}$ are located at a distance of 40 nm on both sides of the quantum well. The process of growth is summarized in Supplementary Fig. 1a.

**Lithography and suspension.** Optical lithography and wet etching are used to pattern large areas, including the boundaries of Hall bars and ohmic contacts. The microstructures used to measure the viscosity and e-e scattering length are formed by electron beam lithography. Reactive ion etching is then performed through resist mask. To suspend the created nanostructures, the sacrificial layer is selectively etched in the HF:$H_2O$ (1:50) solution introduced through 200 nm deep lithographic trenches (Supplementary Fig. 1a).

**Measurements.** Measurements are performed in a dry VTI cryostat in the temperature range from 4 to 50 K in case of magnetic focusing and from 4 to 100 K in case of measurement of superballistic conductance. Experimental samples are equipped with the Au/Ni/Ge ohmic contacts. All electrical measurements are performed using the lock-in technique at the frequency of 70 Hz and the excitation current amplitude of 100 nA (same for injector current in magnetic focusing experiment and current flowing through PC).

**Data availability**

The data that support each plot are available from the corresponding author upon reasonable request.

**Acknowledgements**

The study was funded by the Russian Science Foundation (Grant No. 22-12-00343-П). The nanolithography was performed on the equipment of CKP "VTAN" in ATRC department of NSU.

**Authors contributions**

D.I.S. and D.A.P. suggested, carried the project and analyzed the experimental data. D.A.P. and A.G.P. supervised the project. D.I.S. performed measurements with the technical support of E.Yu.Z. A.A. Shevyrin, A.K.B. and A.A. Shklyaev fabricated samples. D.I.S. and D.A.P. wrote the manuscript. D.A.P., A.G.P., E.Yu.Z. and A.A. Shevyrin contributed to the discussion.

**Competing interests**

Authors declare no conflicts of interests.

**SUPPLEMENTARY INFORMATION**

**Slip electron flow in GaAs microscale constrictions**

D.I. Sarypov *et al.*

**Supplementary Note 1. Device fabrication and transport properties.**

Supplementary Figure 1a shows schematics of the GaAs/AlGaAs heterostructure with a 2DEG, located in the GaAs quantum well. In order to suspend the structures, the sacrificial Al$_{0.8}$Ga$_{0.2}$As layer can be partially removed by wet etching in aqueous HF solution introduced through lithographic trenches. Examples of the suspended structures are shown in Figs. 1b and 1c.

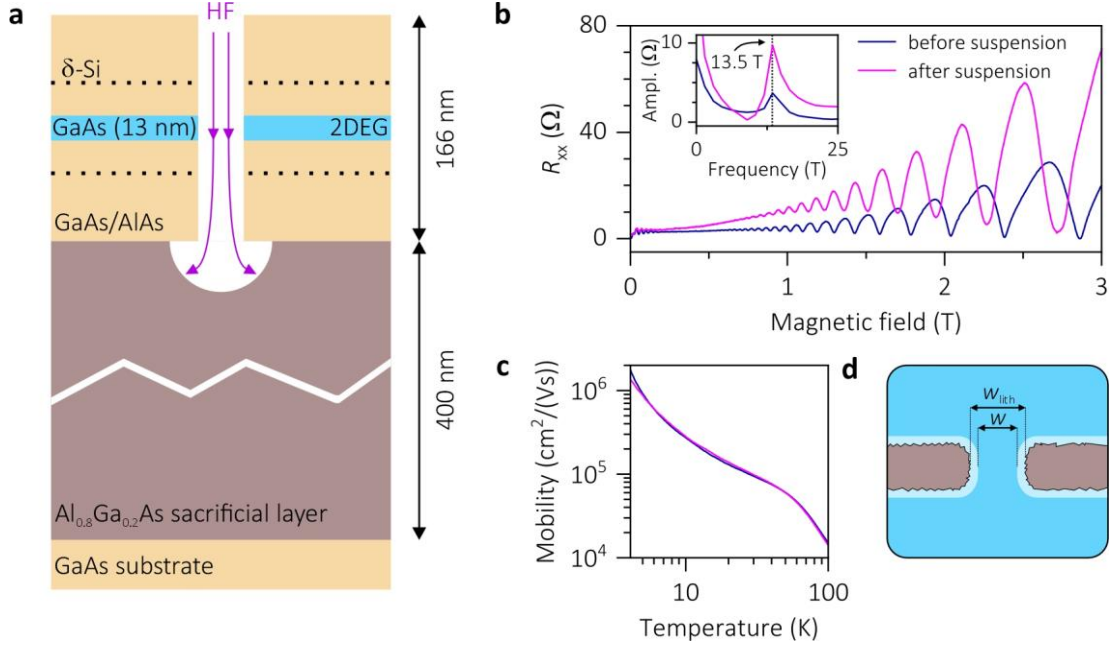

**Supplementary Figure 1| Sample fabrication and transport properties. a,** Schematic view of GaAs/AlGaAs containing 2DEG in GaAs quantum well. **b,** Longitudinal resistance $R_{xx}$, measured at $T = 4$ K as function of magnetic field by the scheme shown in Fig. 3b in main text. Blue (magenta) curve corresponds to measurements performed before (after) the suspension. Inset shows FFT of $R_{xx}$. **c,** Electron mobility extracted from data shown in Fig. 3b with the same color-coding as in **b**. **d,** Schematic image of point contact illustrating the origin of the effective PC width $w$.

Concentration and mobility of the 2DEG are extracted from magnetic-field and temperature dependences of the longitudinal resistance measured by the scheme shown in inset to Fig. 3b. The concentration is determined from Shubnikov–de Haas oscillations (Supplementary Fig. 1b), while the mobility, shown in Supplementary Fig. 1c, is obtained directly from data in Fig. 3b as $R_{2\text{DEG}}(T) = \left(ne\mu(T)\right)^{-1} L/W$ ($L = 10$ μm and $W = 15$ μm are the distance between the voltage probes and the width of the sample, respectively). The electron mobility can further be converted to the momentum relaxation length $l = \mu m v_{\text{F}}/e$, which is shown in Fig. 1d. The values of concentration and mobility provided in main text are the same for non-suspended and suspended samples.

The discussion of electron flow through PC relies on the value of its width $w$. Usually $w$ turns out to be smaller than the lithographic width of the PC $w_{\text{lith}}$[1,2], as illustrated in Supplementary Fig. 1d. In our samples, this is caused by the edge depletion that effectively smoothens the device boundary imperfections. This mechanism establishes the boundary slippage observed in our experiment, where reduced dissipation arises from suppressed momentum transfer at the interfaces.

The edge depletion is usually accounted for by introducing the depletion layer width $w_{\text{depl}}$, which can be determined from the low-$T$ limit of PC resistance. At low temperatures, when electron transport is ballistic, the PC resistance $R$ is described by the Sharvin formula[3], as follows:

$$G_{\text{ball}} = \frac{2e^2}{h}\frac{k_{\text{F}} w}{\pi} = \frac{1}{R_{T=4\text{K}}}, \qquad w = w_{\text{lith}} - w_{\text{depl}}. \qquad (S1)$$

Here we used notation $G_{\text{ball}}$ for ballistic conductance. The depletion layer width, obtained from Eq. (S1), is $w_{\text{depl}} \approx 0.17$ μm before suspension and $w_{\text{depl}} \approx 0.30$ μm after suspension for all PCs. In further analysis we use the $w$ values with $w_{\text{depl}}$ subtracted.

**Supplementary Note 2. Extraction of e-e scattering length from magnetic focusing data.**

As stated in the main text, the theory behind Eq. (2) describes our measurements well. To demonstrate it, suppose first that the actual dependence $l_{\text{ee}}(T)$ is unknown. We can extract it by estimating the values of $R_\infty$ and $R_0$ entering Eq. (1) from the high-*T* and low-*T* limits, respectively.

Thus, at high enough temperatures (50 K in our case), when $l_{\text{ee}} \ll l$, the resonant ballistic trajectory almost completely suppressed by e-e scattering and the height of the magnetic focusing peak is almost zero (Fig. 2c). In this case we can neglect the second term in Eq. (1), so

$$R^*_{T=50\,\text{K}} \approx R_\infty. \tag{S2}$$

At the opposite limit of low temperatures, when $l_{\text{ee}} \gg l$, $\exp(-\alpha l/l_{\text{ee}}) \approx 1$ and Eq. (1) can be written as follows:

$$R^*_{T=4\,\text{K}} \approx R_\infty + R_0 \approx R^*_{T=50\,\text{K}} + R_0. \tag{S3}$$

Combining the above assumptions in Eqs. (S2-S3) with Eq. (1) we obtain:

$$l_{\text{ee}}(T) = \alpha l \frac{1}{\ln\left(\frac{R_0}{R^*(T) - R_\infty}\right)} \approx \alpha l \frac{1}{\ln\left(\frac{R^*_{T=4\,\text{K}} - R^*_{T=50\,\text{K}}}{R^*(T) - R^*_{T=50\,\text{K}}}\right)}. \tag{S4}$$

While this rough estimation is hardly applicable at the extremes of the studied *T*-range, it can be used to compare the measured and theoretical $l_{\text{ee}}$ values at moderate temperatures. According to Eq. (S4), *T*-dependence of $l_{\text{ee}}$ is determined by the measured dependence $R^*(T)$ (shown in Fig. 2c), while its value is limited by the length of the resonant trajectory $l$ and factor $\alpha$. This factor accounts for small-angle scattering effects inherent to the experimental geometry, where finite angular size of detector (the ratio of the detector width to the length of a resonant ballistic trajectory) reduces the impact of e-e scattering on the suppression of ballistic focusing peaks[4,5]. Consequently, the experimentally derived $l_{\text{ee}}$ underestimates its actual value in the 2DEG. We use the factor $\alpha$ as a phenomenological parameter to resolve this mismatch, aligning experimental $l_{\text{ee}}$ values with theoretical predictions from Eq. (2). Namely, we fit the experimental data given by Eq. (S4) with the theory (Supplementary Fig. 2a) and obtain the value $\alpha = 0.57$ which is consistent with previous measurements[6]. Supplementary Fig. 2a shows an agreement with the theory except for the deviations at low and high temperatures caused by the finite difference between $\exp(-\alpha l/l_{\text{ee}})$ and 1 or 0, respectively. Despite these deviations, we conclude that the measured $l_{\text{ee}}$ values agree with the theoretical ones, obtained using Eq. (2).

To correct the deviations, we can fit the experimental data shown in Fig. 2c using Eq. (1), with the known dependence $l_{\text{ee}}(T)$ provided by Eq. (2). As a result of this fitting we get the values of $R_\infty$ and $R_0$ (15.3 Ω and 63.4 Ω) that are close to those estimated by Eqs. (S2-S3) (18 Ω and 58 Ω). The corrected $R_\infty$ and $R_0$ values are then substituted in Eq. (S4) to obtain the refined $l_{\text{ee}}(T)$ dependence, which is shown in Supplementary Fig. 2b. One can see better agreement with the theory than in Supplementary Fig. 2a. The same procedure was carried out before and after the suspension with the same value of $\alpha = 0.57$. In the further analysis given in the main text, we use the $l_{\text{ee}}$ values shown in Supplementary Fig. 2b.

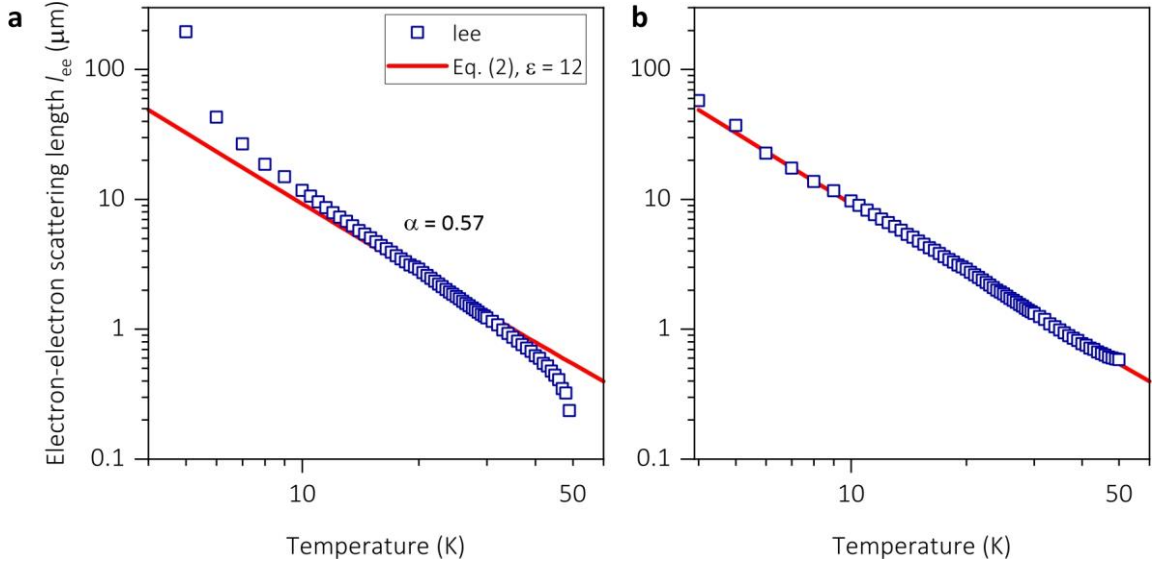

**Supplementary Figure 2| Values of the measured e-e scattering length in non-suspended sample. a,** $l_{ee}(T)$ obtained from Eq. (S4) with $R_\infty$ and $R_0$ estimated from low-$T$ and high-$T$ limits, respectively. Solid line marks the theory provided by Eq. (2). **b,** The same data as on **a**, but with corrected deviations.

**Supplementary Note 3. Quantifying the superballistic conductance.**

The effect of superballistic conductance is observed in PCs of various widths (Supplementary Fig. 3a) located in a single Hall bar (Fig. 1b in the main text). To analyze the $T$-dependences of the PC resistance we follow the common procedure and consider the measured resistance $R$ as the sum of the PC resistance $1/G_{pc}$ and the resistance $R_c$ of the regions adjacent to the PC[7,8]:

$$R = R_c + 1/G_{pc}. \tag{S5}$$

Here the PC conductance $G_{pc} = G_{ball} + G_{vis}$ consists of ballistic $G_{ball}$ and superballistic $G_{vis}$ contributions. The quantity $R_c$ accounts for the momentum-relaxing collisions and leads to the growth of $R$ at high temperature (Supplementary Fig. 3a). We can extract the contribution $R_c$ using an auxiliary sample containing no PC and reproducing the geometry of the regions adjacent to PC (Fig. 1b in main text). Since all momentum-relaxing collisions occur mainly in the regions adjacent to the PC, the contribution $R_c$ and the resistance $R_{2DEG}$ of the auxiliary sample (shown in Fig. 3b in main text) have the same $T$-dependences:

$$R_c(T) = bR_{2DEG}(T).$$

The factor $b$ accounts for the PC boundaries[1] and can be determined experimentally in high-$T$ limit[2]:

$$b = \left(\frac{dR}{dT} \bigg/ \frac{dR_{2DEG}}{dT}\right)\bigg|_{T=100K}.$$

The factor $b$ takes the values of 6–7 for the studied PCs, so, at high temperatures, the quantity $R_c = bR_{2DEG}$ has the same order of magnitude as $R$ (see inset in Supplementary Fig. 3a).

Then we determine the ballistic conductance from the low-$T$ limit of resistance, where $G_{ball} \gg G_{vis}$, and obtain the superballistic conductance from Eq. (S5):

$$G_{vis} = (R - R_c)^{-1} - G_{ball}, \quad G_{ball} = 1/R_{T=4K}$$

Next, we analyze the scaling law of extracted $G_{vis}$ values. Being a signature of electron viscosity, the superballistic conductance $G_{vis}$ is predicted to have distinctive quadratic dependence on the PC width $w$ ($G_{vis} \propto w^2$)[9]. In contrast, the ballistic conductance $G_{ball}$ depends linearly on $w$ (Eq. (S1)). We observe a clear difference in $w$-dependences of $G_{vis}$ and $G_{ball}$ (Supplementary Fig. 3b). The suspension process conserves the observed $w^2$ scaling of $G_{\text{vis}}$ (Supplementary Figs. 3c and 3d), which is consistent with previous measurements[2].

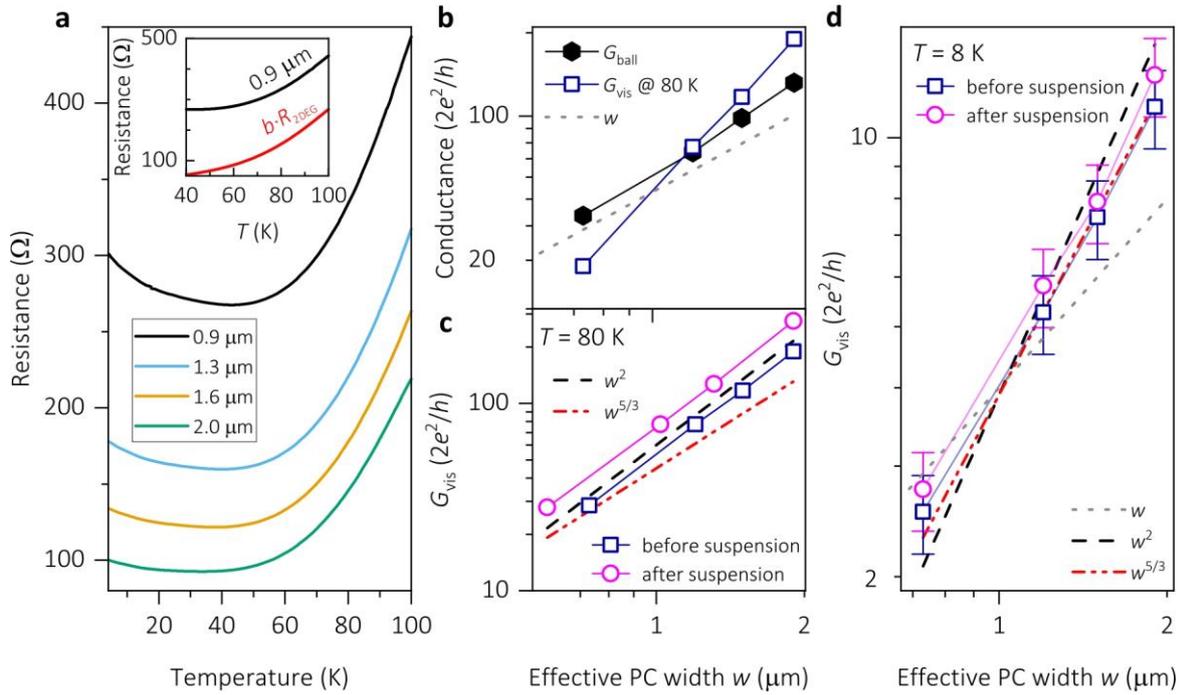

**Supplementary Figure 3| Analysis of superballistic conductance. a**, Resistance of studied PCs as a function of temperature measured before suspension. Numbers correspond to the lithographic width of particular PC. Insert contains the same data for the narrowest PC and extracted contact resistance $R_c = b \cdot R_{\text{2DEG}}$. **b**, Ballistic ($G_{\text{ball}}$) and superballistic ($G_{\text{vis}}$) contributions to the PC conductance as functions of effective PC width $w$ on a log-log scale. The conductance $G_{\text{vis}}(w)$ is plotted at $T = 80$ K. Thin dotted line is a guide for the eye marking the superlinear dependence $G_{\text{vis}}(w)$. **c,d** The dependence $G_{\text{vis}}(w)$ on a log-log scale at $T = 80$ K and $T = 8$ K. Lines are guide for the eye denoting various $w$-dependences. Error bars in **d** are estimated 20 % variation in low-$T$ values of viscosity (Supplementary Note. 4).

**Supplementary Note 4. Precision of the viscosity measurement.**

Our method of extraction of the viscosity values from the measurements of superballistic conductance contains a number of steps (Supplementary Note 3), and each of them may add uncertainty to the final result. To estimate the actual uncertainty, we performed measurements on several identical samples (before and after the suspension) and obtained temperature dependences of viscosity (Supplementary Fig. 4). One can see that temperature dependences, as well as the viscosity reduction after the suspension, are reproduced in all our samples, so the results show good precision.

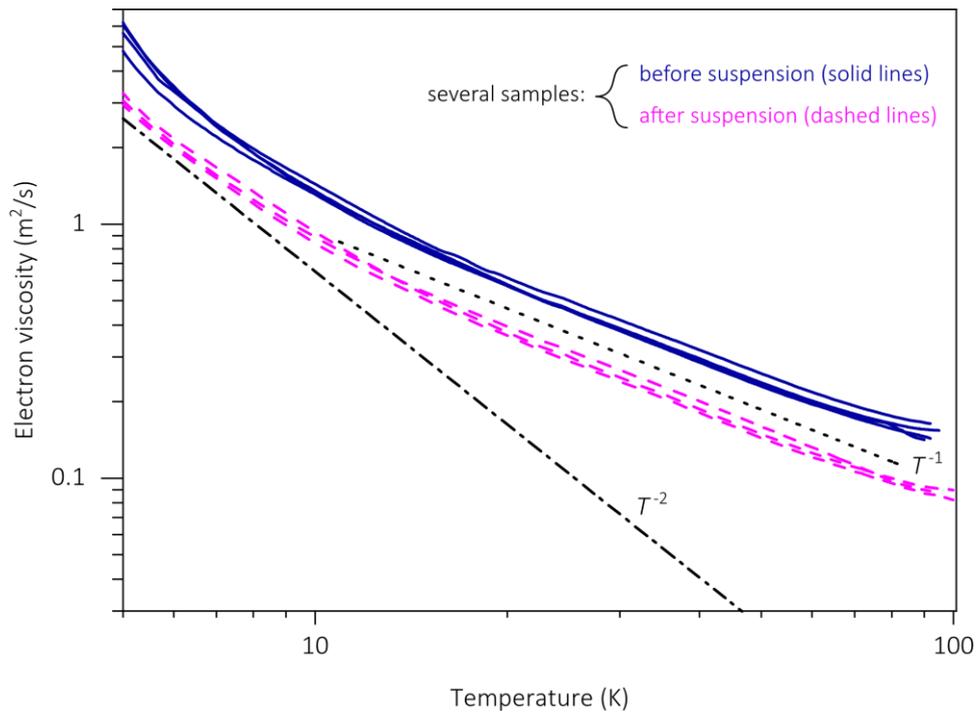

**Supplementary Figure 4| Reproducibility of viscosity measurements.** Electron viscosity obtained by Eq. (3) as a function of temperature, measured on several identical samples.

## Supplementary references